\documentclass[12pt]{article}

\newcommand{\stw}{\sigma^2}

\newcommand{\pmbvstw}{\,^{\pm}\!\mbox{\boldmath$v$}_{\sigma^2}}
\newcommand{\pmRstw}{\,^{\pm}\!R_{\sigma^2}}
\newcommand{\pbvstw}{\,^{+}\!\mbox{\boldmath$v$}_{\sigma^2}}
\newcommand{\pRstw}{\,^{+}\!R_{\sigma^2}}
\newcommand{\mbvstw}{\,^{-}\!\mbox{\boldmath$v$}_{\sigma^2}}
\newcommand{\mRstw}{\,^{-}\!R_{\sigma^2}}
\newcommand{\Rstw}{R_{\sigma^2}}
\newcommand{\Omsth}{\Omega_{\sigma^3}}

\newcommand{\sumstw}{\sum_{\sigma^2}}
\newcommand{\prodsth}{\prod_{\sigma^3}}

\renewcommand{\d}{{\rm d}}
\newcommand{\D}{{\cal D}}
\newcommand{\N}{{\cal N}}
\newcommand{\pN}{\,^{+}\!{\cal N}}
\newcommand{\mN}{\,^{-}\!{\cal N}}

\newcommand{\pmbphi}{\,^{\pm}\!\mbox{\boldmath$\phi$}}
\newcommand{\pmphi}{\,^{\pm}\!\phi}

\newcommand{\pmbv}{\,^{\pm}\!\mbox{\boldmath$v$}}
\newcommand{\pbv}{\,^{+}\!\mbox{\boldmath$v$}}
\newcommand{\mbv}{\,^{-}\!\mbox{\boldmath$v$}}
\newcommand{\bv}{\mbox{\boldmath$v$}}

\newcommand{\bn}{\mbox{\boldmath$n$}}
\newcommand{\bsn}{\mbox{\scriptsize\boldmath$n$}}
\newcommand{\bl}{\mbox{\boldmath$l$}}
\newcommand{\bsl}{\mbox{\scriptsize\boldmath$l$}}
\newcommand{\br}{\mbox{\boldmath$r$}}
\newcommand{\bsr}{\mbox{\scriptsize\boldmath$r$}}
\newcommand{\bvarphi}{\mbox{\boldmath$\varphi$}}
\newcommand{\bpsi}{\mbox{\boldmath$\psi$}}

\newcommand{\pmvstw}{\,^{\pm}\!v_{\sigma^2}}

\newcommand{\vstw}{v_{\sigma^2}}

\newcommand{\pmOm}{\,^{\pm}\!\Omega}
\newcommand{\pOm}{\,^{+}\!\Omega}
\newcommand{\mOm}{\,^{-}\!\Omega}
\newcommand{\pmR}{\,^{\pm}\!R}
\newcommand{\pR}{\,^{+}\!R}
\newcommand{\mR}{\,^{-}\!R}

\newcommand{\dfun}{$\delta$-function }
\newcommand{\dfuns}{$\delta$-functions }

\newcommand{\deth}{\delta^3}

\newcommand{\de}{\delta}

\newcommand{\dektw}{\delta^{(k+2)}}
\newcommand{\dek}{\delta^{(k)}}

\newcommand{\dekmodtw}{\delta^{(k({\rm mod}2)+2)}}

\newcommand{\dekmod}{\delta^{(k({\rm mod}2))}}

\newcommand{\sh}{{\rm sh}}
\newcommand{\ch}{{\rm ch}}

\newcommand{\laonak}{l^{a_1}...l^{a_k}}
\newcommand{\naonak}{n^{a_1}...n^{a_k}}
\newcommand{\daonak}{\partial_{a_1}...\partial_{a_k}}

\newcommand{\ddg}{{{\rm d}\over {\rm d}g}}
\newcommand{\dxdg}{{{\rm d} x\over {\rm d}g}}
\newcommand{\brddgbr}{\left ({{\rm d}\over {\rm d}g} \right )}

\begin{document}

\title{Defining some integrals in Regge calculus}
\author{V.M. Khatsymovsky \\
 {\em Budker Institute of Nuclear Physics} \\ {\em
 Novosibirsk,
 630090,
 Russia}
\\ {\em E-mail address: khatsym@inp.nsk.su}}
\date{}
\maketitle
\begin{abstract}
Regge calculus minisuperspace action in the connection representation has the form in which each term is linear over some field variable (scale of area-type variable with sign). We are interested in the result of performing integration over connections in the path integral. To find this function, we compute its moments, i. e. integrals with powers of that variable. Calculation proceeds through intermediate appearance of $\delta$-functions and integrating them out and leads to finite result for any power. The function of interest should therefore be exponentially suppressed at large areas and it really does being restored from moments. This gives for gravity a way of defining such nonabsolutely convergent integral as path integral.
\end{abstract}


\newpage

Strict definition of the functional integral is possible for Gaussian case; for small deviations from this case it is considered to be definable perturbatively. For general relativity system perturbative expansion is poorly defined due to nonrenormalizability of gravity, and we have non-Gaussian path integral. The action is essentially nonlinear, but in the Cartan-Weyl form, in terms of tetrad and connections, the action can be viewed as linear in some field variable which is bilinear in the tetrad. Suppose we have performed integration over connections and are interested in the dependence of the result on the tetrad bilinears. Of course, different bilinears are not independent, but nothing prevent us from studying analytical properties on the extended region of varying these bilinears as if these were independent variables.

Since we do not possess exact definition of non-Gaussian functional integral, we need its finite dimensional realization on minisuperspace gravity system, or Regge calculus \cite{Regge}. In the connection representation we have
\begin{eqnarray}\label{intDOm}
\N (\pbv, \mbv) = \int \exp {i\over 2} \sumstw \left [ \left (1 + {i\over \gamma} \right ) \sqrt{\pbvstw^2} \arcsin {\pbvstw * \pRstw (\Omega )\over \sqrt{\pbvstw^2} } \right. \nonumber\\ \left. + \left (1 - {i\over \gamma} \right ) \sqrt{\mbvstw^2} \arcsin {\mbvstw * \mRstw (\Omega )\over \sqrt{\mbvstw^2} } \right ] \prodsth \D \Omsth
\end{eqnarray}

\noindent which we have considered, e. g., in recent paper \cite{Kha} and, in more or less similar form, in earlier works. Here $\pmbvstw$ are vectors parameterizing (anti-)selfdual parts $\pmvstw^{ab}$ of the bivector $\vstw^{ab}$ of the triangle $\stw$, $\Omsth$ is connection SO(3,1) matrix on the triangle $\stw$ (holonomy of $\Omega$'s). The $\pmRstw$ is (anti-)selfdual part of $\Rstw$ acting in adjoint, SO(3) representation (to be precise, SO(3,C) matrix). The $\gamma$ is analog of the Barbero-Immirzi parameter \cite{gamma} in our formulation of minisuperspace gravity system. Matrices $\pmOm$, $\pmR$ can be parameterized by complex vector angles $\pmbphi = \bvarphi \mp i\bpsi$ (rotation by the angle $\pmphi = \sqrt{\pmbphi^2}$ around the unit vector $\pmbphi /\sqrt{\pmbphi^2}$).

Consider (\ref{intDOm}) as function of arbitrary $\pbv$, $\mbv = (\pbv)^*$. Consider moments of $\N$ (integral with powers of $\pmbv$). It is convenient to perform continuation to/from the Euclidean-like region. Namely, integration contour over $\bpsi$ considered as complex variable is deformed from real to imaginary values. The $\pmbphi$ become independent real variables, and integration over $\D R$ splits,
\begin{equation}
\D \Omega = \D \pOm \D \mOm, ~~~ \N = \pN \mN, ~~~ \D \pmOm = {\sin^2 (\pmphi/2)\over 4\pi^2 \pmphi^2} \d^3 \pmbphi.
\end{equation}

\noindent We define $\N$ in the region where $\bl$ = $(1 \pm i/\gamma)\pmbv /2$ are real (at $\gamma^{-1} = 0$ this corresponds to real, i. e. timelike $\pmbv$) and then continue to $\pmbv$ of interest. We are faced with integrals of the form
\begin{equation}\label{M}
\int \D \pmR ... \int e^{ilg(\bsn\bsr)} \laonak \d^3 \bl
\end{equation}

\noindent where $l = \sqrt{\bl^2}$, $\bn = \bl / l$, $g(x)$ is odd analytical in the neighborhood of $x = 0$ function, $g(-x) = -g(x)$, such as principal value of $\arcsin x$ of interest or simply $x$. The vector $r^a = \epsilon^a{}_{bc} \pmR^{bc} / 2 = (\pmphi^a \sin \pmphi) / \pmphi$. Existence of integrals (\ref{M}) easily follows at $g(x) = x$: we simply get derivatives of \dfuns $\deth (\br )$ which are then integrated over
\begin{equation}
\D \pmR = \left ({1\over \sqrt{1 - r^2}} - 1 \right ) {\d^3 \br \over 8\pi^2 r^2}.
\end{equation}

\noindent Finiteness is provided by analyticity of this measure at $r \to 0$, $\D \pmR = (c_0 + c_1 r^2 + c_2 (r^2)^2 + ...) \d^3 \br$. Dots in $\int \D R ...$ in (\ref{M}) mean possible dependence on $R$ of factors provided by $\Rstw$ on other triangles $\stw$ due to the Bianchi identities (these are well-behaved for finiteness analytical in $\br$ factors).

For the more general case $g(x) \neq x$ (\ref{M}) is also finite. Again, consideration may go through appearance of \dfuns at an intermediate stage. Namely, special structure of exponential (\ref{M}) allows to extend integration over $l$ to the whole range $( -\infty , +\infty )$. This is only possible because formal substituting $l \to -l$ is equivalent to $\bn \to -\bn$ due to the oddness of $g(x)$. This results in \dfuns of $g$,
\begin{eqnarray}\label{intlg}
 & & \int e^{ilg(\bsn\bsr)} \laonak \d^3 \bl = {1 \over 2} \int \naonak \d^2 \bn \int\limits^{+\infty}_{-\infty} e^{ilg(\bsn\bsr)} l^{k+2} \d l \nonumber\\ & &  \phantom{\int e^{ilg(\bsn\bsr)} \laonak \d^3 \bl} = {1 \over 2} (2\pi) (-i)^{k+2} \int \dektw (g(\bn \br)) \naonak \d^2 \bn.
\end{eqnarray}

\noindent Apply $\dektw (g(x))$ to probe functions,
\begin{equation}\label{probe}
\int \dektw (g(x)) f(x) \d x = \left ( -\ddg \right )^{k+2} \left [ f(x(g)) \dxdg \right ]_{g=0},
\end{equation}

\noindent we find it be combination of $\dektw (x)$, $\dek (x)$, ... , $\dekmodtw (x)$, $\dekmod (x)$. By (\ref{intlg}) read from right to left at $g(x) = x$, leading term $\dektw (\bn \br )$ results in $\int \exp (i \bl \br )\!$ $\!\laonak\!$ $\!\d^3 \bl\!$ $\!= (2\pi)^3 (-i)^k \daonak \deth (\br )$. This being integrated over $\D \pmR$ is finite. Subsequent terms $\de^{(k-2j)} (\bn \br )$, $0 \leq j \leq  [k/2]$ ([k/2] is integer part) result in $(-i)^{k+2} \daonak 2\pi^2\!$ $\!(r^2\!$ $\!+ \varepsilon^2)^{j-1/2} [2^j(2j-1)!!j!]^{-1}$. Here $\varepsilon \to 0$ is introduced to detect possible formation (if indices are contracted) of terms proportional to $\partial^2_a (1/r) = -4\pi \deth (\br )$. Finiteness is provided by analyticity of the measure at $r = 0$,
\begin{equation}
\D \pmR = (P_n (r) + f_n (r)) \theta (1-r) \d^3 \br, ~~~ P_n = \sum^n_{m=0} c_m r^{2m}, ~~~ f_n = O(r^{2n+2}),
\end{equation}

\noindent $\theta (y)$ is Heaviside step function. Convergence of integral with $f_n (r) \theta (1-r)$, $n = [k/2] - j$, is seen immediately, and that of integral with $P_n (r) \theta (1-r)$ follows upon multiple applying integration by parts.

Above we have considered convergence of integrals for moments. If simplified integral is considered,
\begin{equation}\label{N_0}
\N_0 = \int \exp {i\over 2} \left [ \left ( 1 + {i \over \gamma } \right ) \sqrt{\pbv^2} \arcsin { \pbv * \pR \over \sqrt{\pbv^2}} + \left ( 1 - {i \over \gamma } \right ) \sqrt{\mbv^2} \arcsin { \mbv * \mR \over \sqrt{\mbv^2}} \right ] \D R,
\end{equation}

\noindent with no else dependence on $R$ admixed to other triangles $\stw$, calculation of arbitrary moment can be performed in closed form. It is sufficient to integrate in (\ref{M}) and then over $\D \pmR$ the scalars $l^{2k}$. Apply $\de^{(2k+2)} (g(x))$ in (\ref{probe}) to probe functions $f(x) = f^{(2m)} (0) x^{2m} / (2m)!$ and get $(\d / \d g)^{2k+3} [x^{2m+1} (g)] / (2m+1)! |_{g=0}$ for the coefficient of $\de^{(2m)} (x)$ in $\de^{(2k+2)} (g(x))$. Substitute $\de^{(2m)} (x)$ to the RHS of (\ref{intlg}), and this converts to $l^{2m-2}$ in the LHS. Then sum up $l^{2m-2}$ over $m$ with the coefficient found. We have
\begin{eqnarray}\label{moment}
 & & \int \D \pmR \int e^{ilg(\bsn\bsr)} l^{2k} \d^3 \bl \nonumber\\ & & = (-1)^{k+1} \int \D \pmR \int e^{i\bsl\bsr} \d^3 \bl \brddgbr^{2k+3} \left [ \sum^{\infty}_{m=0} (-1)^m {x^{2m+1} (g)\over (2m+1!)} l^{2m-2}\right ]_{g=0} \nonumber\\ & & = (-1)^{k+1} \brddgbr^{2k+2} \left [ \dxdg I(x)\right ]_{g=0}
\end{eqnarray}

\noindent with the "generating function"
\begin{eqnarray}\label{I}
 & & I(x) = \int \left ({1\over \sqrt{1 - r^2}} - 1 \right ) {\d^3 \br \over 8\pi^2 r^2} \int e^{i\bsl\bsr} {\cos (xl) \over l^2} \d^3 \bl \nonumber\\ & & = \pi \int\limits^1_x \left ({1\over \sqrt{1 - r^2}} - 1 \right ) {\d r \over r} = \pi \ln (1 + \sqrt{1 - x^2}).
\end{eqnarray}

\noindent We have extended summation to infinite number of powers of $x$, remembering that upon applying $(\d / \d g)^{2k+3} (\cdot )_{g=0}$ only finite number of terms are active.

In the simplest case $x = g$ the moment (\ref{moment}) is
\begin{equation}
\pi (-1)^{k+1} \brddgbr^{2k+1} \left ({1 \over g} - {1 \over g \sqrt{1 - g^2}}\right )_{g=0}.
\end{equation}

\noindent At $x = \sin g$ the moment is
\begin{equation}
\pi (-1)^{k+1} \brddgbr^{2k+2} \left [ \cos g \ln \left (1 + \cos g\right )\right ]_{g=0}.
\end{equation}

\noindent It is not difficult to find out density of distribution giving these values on monomials (perform a kind of Mellin transform). Appropriate entries in the table of integrals are
\begin{equation}
{1 \over \sqrt{1-g^2}} = {2\over \pi} \int\limits^{\infty}_0 \ch gl K_0(l) \d l,
\end{equation}

\noindent $K_0$ is modified Bessel function, and
\begin{equation}
{g\over 2} \sin g - {1\over 2} + {1\over 2} \cos g \ln [2(1 + \cos g)] = \int\limits^{\infty}_0 {l \over l^2 + 1} {\ch gl\over \sh \pi l} \d l.
\end{equation}

\noindent Thus we find at $g(x) = x$
\begin{equation}\label{my3d94}
\int \D \pmR \int e^{i\bsl \bsr} f ( l^2 ) \d^3 \bl \propto \int {Ki_1(l)\over l}f(-l^2) {\d^3 \bl\over \pi^2}
\end{equation}

\noindent where
\begin{equation}
Ki_1(l) = \int\limits^{\infty}_0 e^{-l\ch \eta} {\d \eta\over \ch \eta} = \int\limits^{\pi /2}_0 \exp \left (-{l\over \sin \varphi}\right ) \d \varphi
\end{equation}

\noindent is integral of $K_0(l) = -[Ki_1(l)]^{\prime}, Ki_1(\infty ) = 0$. This had appeared in our work \cite{Kha2} in 3D SO(3) Regge calculus.

At $g(x) = \arcsin x$ of present interest
\begin{equation}
\int \D \pmR \int e^{ilg(\bsn \bsr)} f ( l^2 ) \d^3 \bl \propto \int {\pi l \over \sh \pi l} {f(-l^2) \over l^2 + 1} {\d^3 \bl \over \pi^2} + (2 \ln 2 - 4) f(1) - 4f^{\prime} (1).
\end{equation}

\noindent The integral in the RHS in both cases is normalized to 1. For the integral (\ref{N_0}) we deform integration contour and pass from $\bl$ to $\bv \equiv \pbv = 2\bl (1 + i/\gamma )^{-1}$ and $\bv^*$,
\begin{equation}
\int \N_0 (v, v^*) f(v^2)h(v^2)^*\d^3 \bv \d^3 \bv^* = \mu (f) \mu (h)^*,
\end{equation}

\noindent where $v = \sqrt{\bv^2}$ and at $\gamma^{-1} = 0$
\begin{equation}
\mu (f) = {1 \over 2} \int {v/2 \over v^2/4 + 1} {f(-v^2) \over \sh (\pi v/2)} \d^3 \bv + \pi (4\ln 2 - 8) f(4) - 32 \pi f^{\prime} (4).
\end{equation}

\noindent Here additional to integral terms correspond to terms in $\N_0 (v, v^*)$ having support at the points $v^2 = 4$. These indicate to interesting possibility to have discrete level in the timelike region $v^2 = 4$. At $\gamma^{-1} \neq 0, \gamma > 0$
\begin{eqnarray}\label{mu-f}
 & & \mu (f) = {i\over 2} \int {(1/\gamma - i) v/2\over (1/\gamma - i)^2 v^2/4 + 1} {f(v^2) \d^3 \bv \over \sh [\pi (1/\gamma - i) v/2]} \nonumber\\ & & \hspace{-1cm} + 4\pi (1 + i/\gamma)^{-3} [(\ln 2 - 2) f(4(1 + i/\gamma)^{-2}) - 8(1 + i/\gamma)^{-2}f^{\prime}(4(1 + i/\gamma)^{-2})].
\end{eqnarray}

\noindent Now non-integral terms have support at nonphysical points $v^2 = 4(1 + i/\gamma)^{-2}$. Thus in any region excluding these points, in particular, in physical region $\Im v^2 = 0$ we have
\begin{equation}\label{Nvv}
\N_0 (v, v^*) = \left | {1 \over {1 \over 4}\left ({1\over \gamma} - i\right )^2 v^2 + 1} \cdot {{1 \over 4}\left ({1\over \gamma} - i\right ) v \over \sh \left [{\pi \over 2} \left ({1\over \gamma} - i\right ) v \right ]} \right |^2.
\end{equation}

Looking at logarithmic expression for generating function (\ref{I}), it is interesting to ask to what extent our definition based on continuation from Euclidean-like region relies on the compactness of SO(4). Alternative way to define the moments of (\ref{N_0}) might proceed via deforming integration contours so that $\bvarphi$ be imaginary. The $\pmbphi$ become independent {\it imaginary} variables (pure Lorentz boost angles), and integration over $\D R$ splits. The $\br$ is imaginary, we define moments on imaginary $\bl$ and find $I(x)$ at imaginary $x$. Logarithmic divergence at $r \to \infty$ appears. Let the cut off for $r$ in (\ref{I}) after $r \to ir$ be $r_0$. Then continue $I(x)$ to real $x$. The $I(x)$ (\ref{I}) is modified by adding constant to logarithm,
\begin{equation}
I(x) \Rightarrow \pi \ln \left ({1 + \sqrt{1 - x^2} \over 1 + \sqrt{1 + r_0^2}}\right ).
\end{equation}

\noindent However, it is not difficult to see that this constant only modifies the coefficient at $f(4(1 + i/\gamma)^{-2})$ in (\ref{mu-f}) and thus does not affect the result (\ref{Nvv}).

In physical spacelike region $\pbv^2 = \mbv^2 = -4|A|^2$, $|A|$ is module of the triangle area, this behaves as $\exp (-2\pi |A|)$ at large $|A|$. In physical timelike region $\pbv^2 = \mbv^2 = 4|A|^2$, this behaves as $\exp (-2\pi |A|/\gamma)$ at large $|A|$. For comparison, replacing $\arcsin x$ by $x$ as in (\ref{my3d94}) gives $\exp (-2 |A|)$ and $\exp (-2 |A|/\gamma)$, respectively.

Above result(s) can be illustrated by the following model integral,
\begin{equation}
\int\limits^{+\infty}_{-\infty} e^{i\sqrt{-v^2} \sh \psi}\d \psi = 2K_0(\sqrt{-v^2}) = \int\limits^{+\infty}_{-\infty} e^{-\sqrt{-v^2} \ch \psi}\d \psi.
\end{equation}

\noindent Here $\sqrt{-v^2}$ is modeling module of the spacelike area, $\psi$ is modeling Lorentz boost angle. This behaves as $\exp (-\sqrt{-v^2})$ at large $v^2$. Nonzero parameter $\gamma^{-1}$ mixes spacelike and timelike area components, so we get $\exp (-\sqrt{v^2}/\gamma)$ for timelike area. Taking integrals over connections (\ref{intDOm}) also reminds calculating Fourier transform of 1, although on curvy (group) manifold. While usual Fourier transform gives \dfun, limiting case of exremely rapidly decreasing exponent, calculations on curvy manifold lead to broadening this \dfun. This looks like what is happening when applying uncertainty principle to conjugate variables. Now conjugate variables are area and connection. Our definition of nonabsolutely convergent integrals thus satisfies such uncertainty principle. In particular, when $g(x) = \arcsin x$, rotation angles enter exponential in the form close to the angles themselves. Then connection integral is more close to the usual Fourier transform and provides larger exponential suppression for areas than the integral at $g(x) = x$ (when exponential contains angles in the more nonlinear manner, as hyperbolic or trigonometric functions).

Appealing feature of definition of integrals through moments adopted is that when computing any moment we only deal with values of finite number of derivatives at $x = 0$, i. e. with local properties of $\arcsin$ function at $x = 0$ at each step. This is important since thus far path integral formalism has been checked in physics only on perturbative level. Nevertheless, knowing full infinite set of moments somehow reproduces nonperturbative feature of this function.

To resume, path integral in the representation of minisuperspace Regge calculus system in terms of rotation matrices in Minkowsky spacetime can be well defined. Upon integrating out connections, contribution of large areas is suppressed exponentially. Thus vertices do not go away to infinity and in this sense the minisuperspace system described by elementary lengths/areas is self-consistent.

The present work was supported in part by the Russian Foundation for Basic Research
through Grant No. 08-02-00960-a.


\end{document}